
\magnification 1200
\hsize=16 true cm
\vsize=23 true cm
\baselineskip=18pt
\vskip 4 cm
\centerline{\bf  A THREE CONFIGURATIONS DIQUARK MODEL FOR BARYONS}
\vskip 2 cm
\centerline { W. S. Carvalho and A. C. B. Antunes}
\vskip 0.5cm
\centerline { UFRJ-Instituto de F\'isica}
\centerline{Caixa postal 68528}
\centerline{21945-970 Rio de janeiro RJ Brasil}
\centerline{ Bitnet: ift10030@UFRJ }

\vskip 3.0cm
\centerline { ABSTRACT }
\vskip 1.0cm
{\it The wave functions in the diquark-quark model for baryons are modified to
take
into account the effect of the different masses of the quarks in diquark
formation.
A spatial wave function is introduced multiplying each flavour-spin
configuration in the quark-diquark model. Futhermore a numerical
coefficient is introduced which weighes the contribution of each
configuration.}



\vfill\eject

\noindent {\bf 1. Introduction }
\vskip 0.5cm

\baselineskip=15pt

The concept of diquark was introduced, in the beginnings of the quark
model to describe baryonic properties. Diquarks may be conceived as
substructures or clusters of pairs of quarks inside baryon wave
function [1,2].
 Diquark-quark models have been intensively developed in many
works on baryon structure and mass spectroscopy, baryon decay and
high energy baryon scattering [3]. Some diquark-quark wave functions
were constructed to describe the structure of baryons. Those
wave functions have the disadvantage of not being completely antisymmetric
under exchange of any two quarks. In fact, they are not antisymmetric under
the exchange of a quark of the diquark and the third quark. This dificulty
is compensated by the simplicity of the model. However, the absence of
antisimmetry is consistent with the diquark interpretation as an
quasielementary object.Futhermore in the baryon wave function construction [4]
mentioned above, it was not taken into account the effect of the difference
in the masses of quarks.

The interactions among the quarks inside the baryons are independent of
the flavours, but the diquark dimension depends on the masses of
the constituent quarks. The higher the quark masses the smaller the dimension
of the diquark.

Given a baryon with three quarks of different masses, there are three different
configurations of diquark-quark that can contribute to its wave function. The
diquarks in each configuration are not equally probable. Another serious
distinction in the three possible configurations of the diquark-quark
model is that their masses are different. This result constitutes an
ambiguity in the diquark-quark model for baryon.

In this paper we give a prescription to construct the baryon wave function,
in which the contribution of each diquark-quark configuration is adequately
taken into account. The expectation value of the Hamiltonian calculated with
the wave function obtained by this prescription gives the baryon mass. The mass
of the baryon is, then, the weighed mean value of the masses of the
diquark-quark configurations. This mean value eliminates the ambiguity in the
computation of the baryon mass. The prescription is applied to the
construction of the wave function of the lowest energy baryonic states in S
wave with spin 1/2. In these wave functions there exists two spin states, S=0
 e S=1, for each S wave diquark. Then each flavour configuration has a scalar
and
a vectorial diquark.

\vskip 0.5 cm

\noindent {\bf 2. Kinematics and Diquark approximation }
\vskip 0.5 cm

The coordinates used in the description of the motion of the three
quarks in the center of mass system of  the baryon are

$$\vec \rho_{i}=\vec r_{j}-\vec r_{k}\eqno(1)$$
and

$$\vec \lambda_{i}=\vec r_{i}-{m_{j}\vec r_{j}+m_{k}\vec r_{k}
\over m_{j}+m_{k}}\eqno(2)$$

The conjugate momenta of these coordinates are

$$\vec p_{i}=\mu_{jk}\dot{\vec \rho_{i}}\eqno(3)$$
and

$$\vec q_{i}=\mu_{i}\dot {\vec \lambda_{i}},\eqno(4)$$
where

$$\mu_{jk}={m_{j}m_{k}\over m_{j}+m_{k}}\eqno(5)$$

$$\mu_{i}={m_{i}\bigl(m_{j}+m_{k})\over m_{1}+m_{2}+m_{3}}
\eqno(6)$$
The kinetic energy of the three quarks is

$$T={{\vec p_{i}}^2\over 2\mu_{jk}}+{{\vec q_{i}}^2\over
2\mu_{i}}\eqno(7)$$

In the diquark model, the potential energy can be written, as
$$V\approx V_{jk}(\vec \rho_{i})+V_{i,jk}(\vec \lambda_{i}).\eqno(8)$$
This is an approximate form of the potential, consistent with the
assumption that diquarks are quasielementary objects inside the baryon.

The Hamiltonian of the baryon, in the diquark model, decouples into
two Hamiltonians,

$$H\approx H_{ij}+H_{i,jk}\eqno(9)$$
The internal motion of the diquark(jk) is described by

$$H_{jk}={{\vec p_{i}^2}\over 2\mu_{jk}}
+V_{jk}(\vec \rho_{i})\eqno(10)$$
and the Hamiltonian of the relative motion of the diquark(jk) and
the third quark(i) is

$$H_{i,jk}={{\vec q_{i}^2}\over 2\mu_{i}}+V_{i,jk}(\vec \lambda_{i})
\eqno(11)$$

The wave function of the three quarks for the i(jk) quark-diquark configuration
factorizes as

$$\psi_{i}(\vec \rho_{i},\vec \lambda_{i})=\phi_{jk}(\vec \rho_{i})\chi
_{i}(\vec \lambda_{i})\eqno(12)$$
Assuming that $\phi_{jk}$ and $\chi_{i}$ are respectively eigenfunctions of

$$H_{jk}\phi_{jk}=E_{jk}\phi_{jk}\eqno(13)$$ and

$$H_{i,jk}\chi_{i}=E_{i,jk}\chi_{i}\eqno(14)$$
The energy of the configuration i(jk) is given by

$$E_{i}=E_{jk}+E_{i,jk}\eqno(15)$$

\noindent {\bf 3. Potentials in diquark model }
\vskip 0.5 cm

The potentials $V_{jk}(\vec \rho_{i})$ and $V_{i,jk}(\vec \lambda_{i})$
have the general form

$$V(\vec r)=V(r)+U_{spin}(\vec r)\eqno(16)$$
The spherically symmetric part of the potential can be written as

$$V(r)=C+V_{v}(r)+V_{s}(r)\eqno(17)$$

$$V_{v}(r)=V_{coul}(r)+(1-f)V_{conf}(r),\eqno(18)$$
where

$$V_{s}(r)=fV_{conf}(r)\eqno(19)$$

$$V_{coul}=-F_{G}{\alpha_{s}\over r}\eqno(20)$$
and

$$V_{conf}=Kr^{1\over 2}\eqno(21)$$
The parameter f controls the proportion of scalar and vectorial
parts under Lorentz transformation of the confining potential [5,6].

For each pair(ij) of particles, $F_{G}$ is given by

$$F_{G}=\bigl<\vec F_{i}\cdot\vec F_{j}\bigr>=F_{ij}^2-F_{i}^2
-F_{j}^2\eqno(22)$$
where $F_{i}^2$ is the expectation value of the quadratic Casimir operator
for the colour group SU(3) in the (i) multiplet.

The parameter C depends on the masses of the interacting particles (ij)
according to

$$C(m_{i},m_{j})=a_{0}+a_{1}x+a_{2}x^2,\eqno(23)$$
where

$$x\equiv ln\bigl(m_{i}^2m_{j}+m_{i}m_{j}^2)\bigr).\eqno(24)$$

The running coupling constant of chromodynamics, $\alpha_{s}$, and the
parameter K of the confining potential are assumed to be constant for
a given colour system and independent of the masses of the interacting
particles.

The potential defined by eqs. (14) to (21) was applied to the
study of the spectroscopy of the mesons $(q\bar q)$ [6], glueballs [7],
and hybrids $(q\bar q g)$ [8]. The masses of the quarks and the parameters
of the potencial used in this work are shown in Table[1].

The spin-dependent terms of the potential can be obtained from the
Breit-Fermi Hamiltonian, [9].

In this paper, only S-wave states are computed. Then the only contributing
spin dependent term is

$$V_{ss}=\bigl({2\over 3m_{1}m_{2}}\bigr)\nabla^2V_{v}(\vec S_{1}\cdot
 \vec S_{2}).\eqno(25)$$

The lowest energy states can be obtained by a variational approach, choosing
gaussian wave functions

$$\phi(r)={\bigl({2\over \pi}\bigr)}^{3/4}{1\over r_{o}^{3/2}}exp\bigl[-
{({r\over r_{o}})}^2\bigr].\eqno(26)$$

The energy and radius $\rho_{0}$ of the diquark(jk) are given by
minimizing

$$E_{jk}\leq\hskip 0.3 cm  <\phi_{jk}\mid H_{jk}\mid \phi_{jk}>.\eqno(27)$$

Similarly, the energy and radius $\lambda_{o}$ of the configuration i(jk)
are obtained by minimizing

$$E_{i,jk}\leq\hskip 0.3 cm  <\chi_{i}\mid H_{i,jk}\mid \chi_{i}>.\eqno(28)$$
In tables (1) to (4), we show the results of these computations.

\vskip 0.5 cm
\noindent {\bf 4. The baryon wavefunctions in quark-diquark model }
\vskip 0.5 cm

Diquarks are assumed to be antitriplet colour states, which
are formed by two colour triplets, the quarks interacting
attractively [10].

The flavour-spin wave functions of the diquarks are denoted by:

$$V_{+1}(q'q)={1\over \sqrt{2}}(qq'+q'q)\otimes
(\uparrow \uparrow)\eqno(29a)$$

$$V_{+1}(q'q)={1\over \sqrt{2}}(qq'+q'q)\otimes{1\over \sqrt{2}}
(\uparrow \downarrow+\downarrow \uparrow)\eqno(29b)$$

$$V_{o}(q'q)={1\over \sqrt{2}}(qq'+q'q)\otimes
(\downarrow \downarrow)\eqno(29c)$$

$$S(q'q)={1\over \sqrt{2}}(qq'-q'q)\otimes{1\over \sqrt{2}}
(\downarrow \uparrow-\uparrow \downarrow)\eqno(29d)$$
where q,q'=(u,d,s,c,b).

The baryon wave functions in the quark-diquark model [4] can be modifyed to
take into account the effect of the difference in quark masses. This
is done by introducing in the baryon wave functions [11]
a factor multiplying each
quark-diquark configuration $q_{a}(q_{b}q_{c})$, which distinguishes
among the contributions of each configuration . These factors are given
by

$$\alpha_{a}={\bigl({\lambda_{a}^s\over \rho_{bc}^s}\bigr)}^
{\nu_{a}},\eqno(30)$$
where (bc) is a scalar diquark, and
$$\beta_{a}={\bigl({\lambda_{a}^v\over \rho_{bc}^v}\bigr)}^
{\nu_{a}},\eqno(31)$$
where (bc) is a vectorial diquark; $\rho_{bc}^s$ and $\rho_{bc}^v$
are the r.m.s radii of the scalar and vectorial(bc) diquarks, respectively.
The r.m.s radii of the configuration a(bc) are $\lambda_{a}^s$
and $\lambda_{a}^v$ for scalar and vectorial diquark (bc), respectively.
The parameters $\nu_{a}$ are fixed by imposing orthogonality on the baryon
wave functions.
The spatial wave functions of a quark-diquark flavour configuration a(bc) will
be denoted by $\phi_{a}^s$ or $\phi_{a}^v$, according as the diquark(bc)
spin is 0 or 1.

The wave functions for baryons in the flavour SU(3) octet,(uds),in the
quark-diquark model,are
given by

$$\mid p,\pm>=\pm F_{p}\biggl\{ \beta_{u}\lbrack V_{0}(ud)u_{\pm}-\sqrt{2}
V_{\pm}(ud)u_{\mp}\rbrack \phi_{u}^v + \beta_{d}\lbrack -\sqrt{2}V_{0}(uu)
d_{\pm}+2V_{\pm}(uu)d_{\mp}\rbrack
\phi_{d}^v + $$
$$\pm 3\alpha_{u}S(ud)u_{\pm}\phi_{u}^s\biggr\}\eqno(32a)$$

$$\mid n,\pm>=\pm F_{n}\biggl\{\beta_{d}\lbrack V_{0}(du)d_{\pm}-\sqrt{2}
V_{\pm}(du)d_{\mp}\rbrack \phi_{d}^v + \beta_{u}\lbrack
-\sqrt{2}V_{0}(dd)u_{\pm}+2V_{\pm}(uu)d_{\mp}\rbrack
\phi_{u}^v + $$
$$ \pm 3\alpha_{d}S(du)d_{\pm}\phi_{d}^s\biggr\}\eqno(32b)$$

$$\mid \Sigma^{+},\pm>=\pm F_{\Sigma^{+}}\biggl\{ \beta_{u}\lbrack V_{0}(us)
u_{\pm}-\sqrt{2}
V_{\pm}(us)u_{\mp}\rbrack \phi_{u}^v + \beta_{s}\lbrack -\sqrt{2}V_{0}(uu)
s_{\pm}+2V_{\pm}(uu)s_{\mp}\rbrack \phi_{s}^v + $$
$$ \pm 3\alpha_{u}S(us)u_{\pm}\phi_{u}^s\biggr\}\eqno(32c)$$

$$\mid \Sigma^{-},\pm>=\pm F_{\Sigma^{-}}\biggl\{ \beta_{d}\lbrack V_{0}(ds)
d_{\pm}-\sqrt{2}
V_{\pm}(ds)d_{\mp}\rbrack \phi_{d}^v + \beta_{s}\lbrack -\sqrt{2}V_{0}(dd)
s_{\pm}+2V_{\pm}(dd)s_{\mp}\rbrack \phi_{s}^v + $$
$$ \pm 3\alpha_{d}S(ds)d_{\pm}\phi_{d}^s\biggr\}\eqno(32d)$$

$$\mid \Sigma^{0},\pm>=\pm F_{\Sigma^{0}}\biggl\{ \beta_{u}\lbrack V_{0}(ds)
u_{\pm}-\sqrt{2}
V_{\pm}(ds)u_{\mp}\rbrack \phi_{u}^v + \beta_{d}\lbrack V_{0}(us)
d_{\pm}-\sqrt{2}V_{\pm}(us)d_{\mp}\rbrack \phi_{d}^v - $$
$$-2\beta_{s}\lbrack V_{0}(ud)s_{\pm}-\sqrt{2}V_{\pm}(ud)s_{\mp}\rbrack
+ \phi_{s}^v
 \pm 3\lbrack \alpha_{d}S(su)d_{\pm}\phi_{d}^s+
 \alpha_{u}S(ds)u_{\pm} \phi_{u}^s\rbrack \biggr\}\eqno(32e)$$

\vskip 2.0cm

$$\mid \Lambda^{0},\pm>=\pm F_{\Lambda^{0}}\biggl\{ \beta_{d}\lbrack -V_{0}(us)
d_{\pm}-\sqrt{2}
V_{\pm}(us)d_{\mp}\rbrack \phi_{d}^v + \beta_{u}\lbrack V_{0}(ds)
u_{\pm}-\sqrt{2}V_{\pm}(ds)u_{\mp}\rbrack \phi_{u}^v + $$
$$ \mp \lbrack
\alpha_{u}S(ds)u_{\pm}\phi_{u}^s+\alpha_{d}S(us)d_{\pm}\phi_{d}^s
-2\alpha_{s}S(du)s_{\pm}\phi_{s}^s\rbrack \biggr\}\eqno(32f)$$

$$\mid \Xi^{0},\pm>=\pm F_{\Xi^{0}}\biggl\{ \beta_{s}\lbrack V_{0}(us)
s_{\pm}-\sqrt{2}
V_{\pm}(us)s_{\mp}\rbrack \phi_{s}^v + \beta_{u}\lbrack -\sqrt{2}V_{0}(ss)
u_{\pm}+2V_{\pm}(ss)u_{\mp}\rbrack \phi_{u}^v + $$
$$ \pm 3\alpha_{s}S(us)s_{\pm}\phi_{s}^s\biggr\}\eqno(32g)$$

$$\mid \Xi^{-},\pm>=\pm F_{\Xi^{-}}\biggl\{ \beta_{s}\lbrack V_{0}(ds)
s_{\pm}-\sqrt{2}
V_{\pm}(ds)s_{\mp}\rbrack \phi_{s}^v + \beta_{d}\lbrack -\sqrt{2}V_{0}(ss)
d_{\pm}+2V_{\pm}(ss)d_{\mp}\rbrack \phi_{d}^v + $$
$$ \pm 3\alpha_{s}S(ds)s_{\pm}\phi_{s}^s\biggr\}\eqno(32h)$$
The wave functions of all baryons denoted by (A,B,C)=(u,d,s,c,b) can be written
in any one of the general forms:

$$\mid AAB,\pm>_{1}=\pm F_{1}\biggl\{ \beta_{A}\lbrack V_{0}(AB)
A_{\pm}-\sqrt{2}
V_{\pm}(AB)A_{\mp}\rbrack \phi_{A}^v$$
$$  +\beta_{B}\lbrack -\sqrt{2}V_{0}(AA)
B_{\pm}+2V_{\pm}(AA)B_{\mp}\rbrack \phi_{B}^v + $$
$$ \pm 3\alpha_{A}S(AB)A_{\pm}\phi_{A}^s\biggr\}\eqno(33a)$$

$$\mid ABC,\pm>_{2}=\pm F_{2}\biggl\{ \beta_{A}\lbrack V_{0}(BC)
A_{\pm}-\sqrt{2}
V_{\pm}(BC)A_{\mp}\rbrack \phi_{A}^v$$
$$  +\beta_{B}\lbrack V_{0}(AC)
B_{\pm}-\sqrt{2} V_{\pm}(AC)B_{\mp}\rbrack \phi_{B}^v + $$
$$-2\beta_{C}\lbrack V_{0}(AB)C_{\pm}
-\sqrt{2}V_{\pm}(AB)C_{\mp}\rbrack \phi_{C}^v$$
$$ \pm 3\lbrack \alpha_{A}S(BC)A_{\pm}\phi_{A}^s+
 \alpha_{B}S(AC)B_{\pm}\bigr)\rbrack \phi_{B}^s\biggr\}\eqno(33b)$$

$$\mid ABC,\pm>_{3}=\pm F_{3}\biggl\{ \beta_{A}\lbrack V_{0}(BC)
A_{\pm}-\sqrt{2}
V_{\pm}(BC)A_{\mp}\rbrack \phi_{A}^v$$
$$ +\beta_{B}\lbrack -V_{0}(AC)
B_{\pm}-\sqrt{2} V_{\pm}(AC)B_{\mp}\rbrack \phi_{B}^v  $$
$$ \mp \lbrack
\alpha_{A}S(CB)A_{\pm}\phi_{A}^s+\alpha_{B}S(CA)B_{\pm}\phi_{B}^s
-2\alpha_{C}S(AB)C_{\pm}\phi_{C}^s\rbrack \biggr\}.\eqno(33c)$$
The normalization constants are

$$F_{1}=\bigl(9\alpha_{A}^2+3\beta_{A}^2+6\beta_{B}^2\bigr)^{-1/2}\eqno(34a)$$

$$F_{2}=\biggl(9\bigl(\alpha_{A}^2+\alpha_{B}^2\bigr)+
3\bigl(\beta_{A}^2+\beta_{B}^2+4\beta_{C}^2\bigr)\biggr)^{-1/2}\eqno(34b)$$

$$F_{3}=\biggl(\alpha_{A}^2+\alpha_{B}^2+4\alpha_{C}^2+3\bigl(\beta_{A}^2
+\beta_{B}^2\bigr)\biggr)^{-1/2}\eqno(34c)$$

The orthogonality conditions, which determine the values of the parameters
$\nu_{a}$, are

$$_{2}<ABC,\pm \mid ABC,\pm>_{3}=0,\eqno(35)$$
which gives

$$(\beta_{A}^2-\beta_{B}^2)-(\alpha_{A}^2-\alpha_{B}^2)=0\eqno(36)$$

The values of the parameters $\nu_{a}$ satisfying these conditions are given
in Table [2]. The values of $\nu_{a}$ for states like $\mid AAB,\pm>_{1}$
are undetermined by these conditions, and are assumed to be $\nu_{a}=0.6$,
a mean value in the interval [0.2,1.0] that contains the values of
$\nu_{a}$ shown in Table [2].Choosing $\nu_{a}=0.6$, we have a general value
that givesthe lowest values for the masses of states like $\mid AAB>_{1}$,
which is consistent with a linear variation function approach for baryon
wave functions.

Using the above defined notations (32), the wave functions of the spin 1/2
charmed baryons pertaining to 20-plet of the flavour SU(4) can be written
as in Table [3]. Similarly, the wave functions of the spin 1/2 baryons with
bottom flavour are shown in Table [4].
The values of the factors $\alpha$ and $\beta$ defined in eqs (30,31),
multiplying
each flavour configuration a(bc), are listed in Tables [5] and [6].

The differences in the values of the
$\alpha_{a}$,$\alpha_{b}$,$\alpha_{c}$ and
$\beta_{a}$,$\beta_{b}$,$\beta_{c}$ for a baryon with flavours(a,b,c)
show that the formation of the diquarks (bc),(ac) and (ab) are not
equally probable inside the baryon. The most probable configuration
is the one in which the lighest quark is outside the diquark.

The mass of the baryon with quarks(A,B,C) is given by

$$M_{(A,B,C)}=m_{a}+m_{b}+m_{c}+<ABC\mid H \mid ABC>.\eqno(37)$$

The expectation values of the Hamiltonian for each one of the states (33) are

$$E_{(AAB)_{1}}=F_{1}^2\bigl(3\beta_{A}^2E_{(AB)A}^v+6\beta_{B}^2E_{(AA)B}^v
+9\alpha_{A}^2E_{(AB)A}^s\bigr)\eqno(38a)$$

$$E_{(ABC)_{2}}=F_{2}^2\biggl(3\beta_{A}^2E_{(BC)A}^v+3\beta_{B}^2E_{(AC)B}^v
+12\beta_{C}^2E_{(AB)C}^v+9\bigl(\alpha_{A}^2E_{(BC)A}^s+\alpha_{B}^2
E_{(AC)B}^s\bigr)\biggr)\eqno(38b)$$

$$E_{(ABC)_{3}}=F_{3}^2\biggl(3\beta_{A}^2E_{(BC)A}^V+3\beta_{B}^2E_{(AC)B}^v
+\alpha_{A}^2E_{(BC)A}^s+\alpha_{B}^2E_{(BC)A}^s+4\alpha_{C}^2E_{(AB)C}^s
\biggr)\eqno(38c)$$

The masses of the spin 1/2 S-wave baryons obtained by our method are shown
in Tables [7] and [8], and compared with the existing experimental values.

\vskip 0.3cm

\noindent {\bf Acknowledgments}

The authors are grateful to Dr. Helio Freitas de Carvalho and Dr. Antonio
Soares de
Castro for many stimulating discussions. This work was partially supported by
CNPq and FINEP.
$$\centerline{Table [1]-values of parameters}$$
$$\vbox{\halign{\hfil#\hfil\cr
\noalign{\hrule}
\ \cr
$ m_u = m_d = 0.38GeV, m_s = 0.5GeV, m_c = 1.5GeV, m_b = 4.5GeV$ \cr
$ a_0 = 0.0010, a_1 = 0.146, a_2 = -1.412$ \cr
$ f=0.5, \alpha_s = 0.187, K = 0.767$ \cr
$ F_G = -2/3 (diquark), F_G = -4/3 (meson)$ \cr
\ \cr
\noalign{\hrule}}}$$

$$\centerline{Table [2]-Values of the parameters $\nu_{a}$, for the}$$
$$\centerline{quark-diquark configurations a(bc). Where (q=u,d).}$$
$$\vbox{\halign{\hfil#\hfil& \qquad\hfil
#\hfil&\qquad\hfil#\hfil&\qquad\hfil#\hfil&\qquad\hfil#\hfil\cr
\noalign{\hrule}
\ \cr
Bar. & $\nu_{q}$ & $\nu_{s}$ & $\nu_{c}$ & $\nu_{b}$\cr
\ \cr
\noalign{\hrule}
\ \cr
qsc & 0.6957 & 0.9824 & 0.5 & -\cr
qsb & 0.3934 & 0.5    & -   & 0.4443\cr
qcb & 0.7811 & -      & 0.5 & 0.9823\cr
scb &   -    & 0.3472 & 0.5 & 0.2493\cr
\ \cr
\noalign{\hrule}}}$$

\vskip 5cm

$$\centerline{Table[3]-Wave functions, in the quark-diquark model, for}$$
$$\centerline{baryons with charm flavour, in the spin 1/2 S wave.}$$
$$\vbox{\halign{\hfil#\hfil\cr
\noalign{\hrule}
\ \cr
Wave Functions with charm quark\cr
\ \cr
\noalign{\hrule}
\ \cr
$\mid \Sigma_{c}^{++},\pm>\hskip 0.3cm =\hskip 0.3cm \mid uuc,\pm>_{1}$  ;
 $\mid \Sigma_{c}^{+},\pm>\hskip 0.3cm = \hskip 0.3cm\mid udc,\pm>_{2}$ \cr
$\mid \Sigma_{c}^{0},\pm>\hskip 0.3cm =\hskip 0.3cm \mid ddc,\pm>_{1}$  ;
 $\mid \Lambda_{c}^{+},\pm>\hskip 0.3cm = \hskip 0.3cm\mid udc,\pm>_{3}$ \cr
$\mid \Xi_{c}^{+},\pm>\hskip 0.3cm =\hskip 0.3cm \mid usc,\pm>_{2}$  ;
 $\mid \Xi_{c}^{+*},\pm>\hskip 0.3cm = \hskip 0.3cm\mid usc,\pm>_{3}$ \cr
$\mid \Xi_{c}^{0},\pm>\hskip 0.3cm =\hskip 0.3cm \mid dsc,\pm>_{2}$  ;
 $\mid \Xi_{c}^{0*},\pm>\hskip 0.3cm = \hskip 0.3cm\mid dsc,\pm>_{3}$ \cr
$\mid \Omega_{c}^{0},\pm>\hskip 0.3cm =\hskip 0.3cm \mid ssc,\pm>_{1}$  ;
 $\mid \Omega_{cc}^{+},\pm>\hskip 0.3cm = \hskip 0.3cm\mid ccs,\pm>_{1}$ \cr
$\mid \Xi_{cc}^{++},\pm>\hskip 0.3cm =\hskip 0.3cm \mid ccu,\pm>_{1}$  ;
 $\mid \Xi_{cc}^{+},\pm>\hskip 0.3cm = \hskip 0.3cm\mid ccd,\pm>_{1}$ \cr
\ \cr
\noalign{\hrule}}}$$

\vskip 3cm

$$\centerline{Table[4]-Wave functions, in the quark-diquark model, for}$$
$$\centerline{baryons with bottom flavour, in the spin 1/2 S wave.}$$
$$\vbox{\halign{\hfil#\hfil\cr
\noalign{\hrule}
\ \cr
Wave Functions with bottom quark\cr
\ \cr
\noalign{\hrule}
\ \cr
$\mid \Sigma_{b}^{+},\pm>\hskip 0.3cm =\hskip 0.3cm \mid uub,\pm>_{1}$  ;
 $\mid \Sigma_{b}^{-},\pm>\hskip 0.3cm = \hskip 0.3cm\mid udb,\pm>_{2}$ \cr
$\mid \Sigma_{b}^{-},\pm>\hskip 0.3cm =\hskip 0.3cm \mid ddb,\pm>_{1}$  ;
 $\mid \Lambda_{b}^{0},\pm>\hskip 0.3cm = \hskip 0.3cm\mid udb,\pm>_{3}$ \cr
$\mid \Xi_{b}^{0},\pm>\hskip 0.3cm =\hskip 0.3cm \mid usb,\pm>_{2}$  ;
 $\mid \Xi_{b}^{0*},\pm>\hskip 0.3cm = \hskip 0.3cm\mid usb,\pm>_{3}$ \cr
$\mid \Xi_{b}^{-},\pm>\hskip 0.3cm =\hskip 0.3cm \mid dsb,\pm>_{2}$  ;
 $\mid \Xi_{c}^{-*},\pm>\hskip 0.3cm = \hskip 0.3cm\mid dsb,\pm>_{3}$ \cr
$\mid \Omega_{b}^{-},\pm>\hskip 0.3cm =\hskip 0.3cm \mid ssb,\pm>_{1}$  ;
 $\mid \Omega_{bb}^{-},\pm>\hskip 0.3cm = \hskip 0.3cm\mid sbb,\pm>_{1}$ \cr
$\mid \Xi_{bb}^{+},\pm>\hskip 0.3cm =\hskip 0.3cm \mid bbu,\pm>_{1}$  ;
 $\mid \Xi_{bb}^{-},\pm>\hskip 0.3cm = \hskip 0.3cm\mid dbb,\pm>_{1}$ \cr
$\mid \Xi_{cb}^{+},\pm>\hskip 0.3cm =\hskip 0.3cm \mid ucb,\pm>_{2}$  ;
 $\mid \Xi_{cb}^{+*},\pm>\hskip 0.3cm = \hskip 0.3cm\mid ucb,\pm>_{3}$ \cr
$\mid \Xi_{cb}^{0},\pm>\hskip 0.3cm =\hskip 0.3cm \mid dcb,\pm>_{2}$  ;
 $\mid \Xi_{cb}^{0*},\pm>\hskip 0.3cm = \hskip 0.3cm\mid dcb,\pm>_{3}$ \cr
$\mid \Xi_{scb}^{0},\pm>\hskip 0.3cm =\hskip 0.3cm \mid scb,\pm>_{2}$  ;
 $\mid \Xi_{sbb}^{0*},\pm>\hskip 0.3cm = \hskip 0.3cm\mid scb,\pm>_{3}$ \cr
$\mid \Omega_{bbc}^{0},\pm>\hskip 0.3cm =\hskip 0.3cm \mid bbc,\pm>_{1}$  ;
 $\mid \Omega_{ccb}^{+},\pm>\hskip 0.3cm = \hskip 0.3cm\mid ccb,\pm>_{1}$ \cr
\ \cr
\noalign{\hrule}}}$$

\vskip 5cm

$$\centerline{Table [5]-Values of the factors $\alpha_{a}$ for configuration}$$
$$\centerline{a(bc) with scalar diquark (bc). Where (q=u,d).}$$
$$\vbox{\halign{\hfil#\hfil& \qquad\hfil
#\hfil&\qquad\hfil#\hfil&\qquad\hfil#\hfil&\qquad\hfil#\hfil\cr
\noalign{\hrule}
\ \cr
diquark & $\alpha_{q}$ & $\alpha_{s}$ & $\alpha_{c}$ & $\alpha_{b}$\cr
\ \cr
\noalign{\hrule}
\ \cr
qq & 0.8443 & 0.7818 & 0.6891 & 0.6468\cr
qs & 0.8508 & 0.7860 & 0.7263 & 0.7127\cr
ss & 0.8664 & -      & 0.6855 & 0.6301\cr
qc & 0.8664 & 0.6779 & 0.6394 & 0.3824\cr
sc & 0.8655 & 0.8571 & 0.6713 & 0.7979\cr
cc & 1.0358 & 0.9825 & -      & 0.6438\cr
qb & 0.8627 & 0.8153 & 0.6550 & 0.4809\cr
sb & 0.9290 & 0.8655 & 0.6909 & 0.5119\cr
cb & 1.1864 & 1.0369 & 0.8412 & 0.6394\cr
bb & 1.3091 & 1.2346 & 0.9754 & -\cr
\ \cr
\noalign{\hrule}}}$$

$$\centerline{Table [6]-Values of the factors $\beta_{a}$ for configuration}$$
$$\centerline{a(bc) with scalar diquark (bc). Where (q=u,d).}$$
$$\vbox{\halign{\hfil#\hfil& \qquad\hfil
#\hfil&\qquad\hfil#\hfil&\qquad\hfil#\hfil&\qquad\hfil#\hfil\cr
\noalign{\hrule}
\ \cr
diquark & $\beta_{q}$ & $\beta_{s}$ & $\beta_{c}$ & $\beta_{b}$\cr
\ \cr
\noalign{\hrule}
\ \cr
qq & 0.4411 & 0.6776 & 0.5927 & 0.5601\cr
qs & 0.7719 & 0.6966 & 0.6542 & 0.6531\cr
ss & 0.7964 & -      & 0.6152 & 0.5702\cr
qc & 0.8117 & 0.5980 & 0.5853 & 0.3375\cr
sc & 0.8047 & 0.8114 & 0.6224 & 0.7725\cr
cc & 1.001 &  0.9477 & -      & 0.6111\cr
qb & 0.8217 & 0.7758 & 0.5943 & 0.3937\cr
sb & 0.9011 & 0.8249 & 0.6448 & 0.4457\cr
cb & 1.1454 & 1.0199 & 0.8145 & 0.6061\cr
bb & 1.2861 & 1.2122 & 0.9530 & -\cr
\ \cr
\noalign{\hrule}}}$$

\raggedbottom

\vskip 14cm

$$\centerline{Table [7]-Masses (in GeV) of S wave and spin 1/2 baryons}$$
$$\centerline{with experimental datas [12].}$$
$$\vbox{\halign{\hfil#\hfil& \qquad\hfil
#\hfil&\qquad\hfil#\hfil&\qquad\hfil#\hfil\cr
\noalign{\hrule}
\ \cr
Flavour & Baryon & Our Results & Experimental\cr
\ \cr
\noalign{\hrule}
\ \cr
uud & p & 1.0402 & 0.9383\cr
udd & n & 1.0402 & 0.9396\cr
uus & $\Sigma^{+}$ & 1.2064 & 1.1894\cr
uds & $\Sigma^{0}$ & 1.2064 & 1.1974\cr
uds & $\Lambda^{0}$ & 1.1823 & 1.1156\cr
dds & $\Sigma^{-}$ & 1.2064 & 1.1926\cr
uss & $\Xi^{0}$ & 1.3381 & 1.3149\cr
dss & $\Xi^{-}$ & 1.3381 & 1.3213\cr
uuc & $\Sigma_{c}^{++}$ & 2.5193 & 2.453\cr
udc & $\Sigma_{c}^{+}$ & 2.5193 & 2.453$\pm$ 0.003\cr
udc & $\Lambda_{c}^{+}$ & 2.4296 & 2.285\cr
ddc & $\Sigma_{c}^{0}$ & 2.5193 & 2.453\cr
udb & $\Sigma_{b}^{0}$ & 5.8639 & 5.641$\pm$ 0.050\cr
\ \cr
\noalign{\hrule}}}$$

\vskip 9cm

$$\centerline{Table [8]-Masses (in GeV) of S wave and spin }$$
$$\centerline{1/2 baryons without experimental datas.}$$
$$\vbox{\halign{\hfil#\hfil& \qquad\hfil
#\hfil&\qquad\hfil#\hfil\cr
\noalign{\hrule}
\ \cr
Flavour & Baryon & Our Results\cr
\ \cr
\noalign{\hrule}
\ \cr
usc & $\Xi_{c}^+,\Xi_{c}^{+*}$ & 2.6534,2.5843\cr
dsc & $\Xi_{c}^0,\Xi_{c}^{0*}$ & 2,6534,2.5843\cr
ssc & $\Omega_{c}^{0}$ & 2.7536\cr
ucc & $\Xi_{cc}^{++}$ & 3.7580\cr
dcc & $\Xi_{cc}^{+}$ & 3.7580\cr
ssc & $\Omega_{cc}^{+}$ & 3.8607\cr
uub & $\Sigma_{b}^{+}$ & 5.9862\cr
ddb & $\Sigma_{b}^{-}$ & 5.9862\cr
usb & $\Xi_{b}^{0},\Xi_{b}^{0*}$ & 6.1154,6.0214\cr
dsb & $\Xi_{b}^{-},\Xi_{b}^{-*}$ & 6.1154,6.0214\cr
ssb & $\Omega_{b}^-$ & 6.1948\cr
ucb & $\Xi_{cb}^{+},\Xi_{cb}^{+*}$ & 7.0949,7.1241\cr
dcb & $\Xi_{cb}^{0},\Xi_{cb}^{0*}$ & 7.0949,7.1241\cr
scb & $\Xi_{scb}^{0},\Xi_{scb}^{0*}$ & 7.1719,7.1997\cr
ccb & $\Omega_{ccb}^{+}$ & 8.2060\cr
ubb & $\Xi_{bb}^{0}$ & 10.3836\cr
dbb & $\Xi_{bb}^{-}$ & 10.3836\cr
sbb & $\Omega_{bb}^{-}$ & 10.4961\cr
cbb & $\Omega_{bbc}^{0}$ & 10.4961\cr

\ \cr
\noalign{\hrule}}}$$

\noindent{\bf References:}

\item{[1.]}T. Kobayashi and N. Namik, Prog. in Theor. Phys.,
{\bf 33} (1965).
\item{[2a.]}D.B. Lichtenberg et al., Z. Phys. C-
Particles and Fields{\bf 19} (1983) 19.
\item{[2b.]}D.B. Lichtenberg et al., Phys. Rev. {\bf 155} (1973) 1601.
\item{[3.]}B.O. Skytt and S. Fredrisson, {\it Diquark-91 a
literature Survey, Lulea University of Technology Report}.
\item{[4.]}M. Anselmino,F. Caruso, S. Forte and B. Pire,
Phys. Rev {\bf D38} 11, (1988) 3516.
\item{[5.]}H.F. Carvalho et al., Lett. Nuovo Cim.
{\bf 34},6 (1978).
\item{[6.]}A.S. Castro,H.F. Carvalho and A.B. d'Oliveira,
Lett. Nuovo Cim. {\bf 43} 161 (1985).
\item{[7.]}A.S. Castro,H.F. Carvalho and A.C.B. Antunes,
Nuovo Cim. A101 (1989) 423.
\item{[8.]} IF/UFRJ/93/04.
\item{[9.]}W. Lucha,F.F. Schoberl and D. Gromes,
Phys. Rep. {\bf 200} 4 (1991).

\item{[10.]}H.J. Lipkin, Phys. Lett. {\bf 74B} 4,5 (1978) 399.
\item{[11.]}V. L. Cherniak et al., Nucl. Phys. B {\bf 246} (1984) 143.
\item{[12.]}Phys. Rev. {\bf D45} "{\it Particles and Fields}"
11-II (1992).

\bye